\documentclass[twocolumn,superscriptaddress,showpacs,nofootinbib,notitlepage,preprintnumbers,secnumarabic,amssymb, nobibnotes, aps, prl]{revtex4-2}
\usepackage{graphicx}
\usepackage{epstopdf}
\usepackage{amsmath}
\usepackage{amsfonts}
\usepackage{amssymb}
\usepackage{appendix}
\usepackage{enumerate}
\usepackage{natbib}
\usepackage{comment}
\usepackage{bbold}
\usepackage[shortlabels]{enumitem}
\usepackage{color}
\usepackage{slashed}
\usepackage{subfigure}
\usepackage{setspace}
\usepackage{footnote}
\usepackage{lipsum}
\usepackage{multirow}
\usepackage[colorlinks = true,
            linkcolor = blue,
            urlcolor  = blue,
            citecolor = blue,
            anchorcolor = blue]{hyperref}
\usepackage[capitalize]{cleveref}
\usepackage{braket}
\usepackage[compat=1.1.0]{tikz-feynman}
\usepackage{multirow}
\usepackage{physics}
\usepackage{feynmp-auto}
\usepackage[normalem]{ulem}
\usepackage{url}
\usepackage{units}

\def\bar{\overline}

\begin{document}

\singlespacing

\preprint{USTC-ICTS/PCFT-22-30}

\title{Neutrino Origin of LHAASO's 18 TeV GRB221009A Photon }

\author{Vedran Brdar} 
\email{vedran.brdar@cern.ch}
\affiliation{Theoretical Physics Department, CERN,
Esplande des Particules, 1211 Geneva 23, Switzerland}
\author{Ying-Ying Li}
\email{yingyingli1013@outlook.com}
\affiliation{Interdisciplinary Center for Theoretical Study,
University of Science and Technology of China, Hefei, Anhui 230026, China
}
\affiliation{Peng Huanwu Center for Fundamental Theory, Hefei, Anhui 230026, China}

\begin{abstract}
LHAASO collaboration detected photons with energy above 10 TeV from the most recent gamma-ray burst (GRB), GRB221009A. Given the redshift of this event, $z\sim 0.15$, photons of such energy are expected to interact with the diffuse extragalactic background light (EBL) well before reaching Earth. In this paper we provide a novel neutrino-related explanation of the most energetic 18 TeV event reported by LHAASO. We find that the minimal viable scenario involves both mixing and transition magnetic moment portal between light and sterile neutrinos. The production of sterile neutrinos occurs efficiently via mixing while the transition magnetic moment portal governs the decay rate in the parameter space where tree-level decays via mixing to non-photon final states are suppressed. Our explanation of this event,  while being consistent with the terrestrial constraints, points to the non-standard cosmology.    
\end{abstract}

\maketitle

\textbf{Introduction.}
The detection of recent GRB221009A event at $z \simeq 0.15$ \cite{redshift}  was first reported by \textit{Swift}-Burst Alert Telescope \cite{SBAT} and the \textit{Fermi} Gamma-ray burst Monitor (GBM) \cite{GBM}. It was followed by an array of complementary observations among which those  by LHAASO \cite{LHAASO} and CARPET-2 \cite{Carpet} stand out due to very high-energetic photons reported in association with this GRB event. LHAASO detected around 5000 events across $\sim 2000$ sec window and the energy of the most energetic photon is around 18 TeV. It is known that the interaction rate of such high energetic extragalactic photons is determined chiefly by the cross section for scattering on the extragalactic background light (EBL) \cite{Gould:1966pza,ebl2}; it turns out that the likelihood for propagation of 18 TeV photon from $z\sim 0.15$ to Earth without scattering is around $e^{-15}\sim 10^{-7}$ \cite{Dom_nguez_2010, Franceschini:2017iwq}.  Such a small probability motivated more than several studies including attempts to solve this puzzle with physics beyond the Standard Model. The idea is to have a weakly interacting particle propagating most of the distance between the GRB source and Earth; clearly such particles need to have certain interactions with photons. Discussions related to axion-like particles \cite{Troitsky:2022xso,Lin:2022ocj,Nakagawa:2022wwm,Zhang:2022zbm} and sterile neutrinos appeared \cite{Smirnov:2022suv,Cheung:2022luv}, and for completeness let us also point out studies related to Lorentz invariance violation \cite{Finke:2022swf}.

Here we would already like to stress that when comparing GRB221009A photon flux observed by Fermi-LAT with LHAASO's observation of high-energy events, attenuation of $\approx 10^{-7}$ is in the ballpark of values that can reasonably well explain both datasets \cite{Finke:2022swf}. However, for LHAASO's observation itself, single 18 TeV event should imply couple of orders of magnitude more than the reported $\mathcal{O}(5000)$ events at lower energies.
 
As far as the 251 TeV photon observed by CARPET-2 is concerned, while it is reported  in time-correlation with GRB221009A, its orgin from this GRB source would imply the rise of the flux above 100 TeV energies and explanations from both standard and non-standard physics perspective appears unfeasible \cite{Gonzalez:2022opy}. This event most likely stems from another source previously identified by HAWC \cite{Hawc} and the time-correlation with GRB221009A is likely only a coincidence. \\

 
\textbf{Motivation for sterile neutrino. The Model.}
In addition to photons, neutrinos are also produced in GRB events. IceCube collaboration has previously performed dedicated searches \cite{IceCube:2014jkq,IceCube:2022rlk} in order to possibly associate some of the GRB events with $\mathcal{O}(100\,\text{TeV}-1\,\text{PeV})$ diffuse extragalactic neutrino background \cite{IceCube:2014stg}. IceCube has also constrained the time-integrated flux of neutrinos from GRB221009A at Earth which also allows us to learn more about the sources themselves \cite{Murase:2022vqf}.
Non-observation of track-like events from GRB221009A converts to $E^2 dN_\nu/ dE < 3.9\times 10^{-2} \unit[]{GeV\times cm^{-2}}$ at 90\% CL \cite{icecube}. 

In this work we argue the possibility that 18 TeV photon stems from neutrinos produced at the source. Namely, if sterile neutrinos exist, they can simply be produced in association with left-handed neutrinos. Then, after propagating large distance, sterile neutrinos would decay to photons, ideally at locations where EBL would not significantly attenuate this secondary photon flux. In the context of LHAASO's 18 TeV event, this idea was recently discussed in two papers which considered transition magnetic moment portal and mixing between active and sterile neutrinos, respectively. While for the former case, discussed in \cite{Cheung:2022luv}, the sterile neutrino flux is heavily suppressed by a factor proportional to the magnetic moment which is severely constrained, the hindrance for the mixing case \cite{Smirnov:2022suv} is the relative suppression of the photon flux due to the small branching ratio for sterile neutrino decay into photon.

In what follows we introduce a scenario featuring both transition magnetic moment portal and mixing between active and sterile neutrinos. Given that IceCube searched for track-like events, we will work with the assumption that the only non-vanishing mixing between active and sterile neutrinos is in the muon flavor; we denote the respective matrix element of the extended $4\times 4$ leptonic mixing matrix with $U_{\mu 4}$. The relevant part of the Lagrangian reads
\begin{align}
\mathcal{L}\supset d_\mu \bar{\nu_{\mu L}} \sigma_{\mu\nu} F^{\mu\nu} N + \frac{1}{2} m_N \bar{N}^c N + y \bar{L_\mu} \tilde{H} N + \text{h.c.}\,.
\label{eq:1}
\end{align}
We choose the transition magnetic moment portal to exist only between muon and sterile neutrino, $N$, and its strength is $d_\mu$ in the unit of inverse energy (since we are not after active neutrinos from sterile neutrino decay, flavor universal assumption would work equally well). In \cref{eq:1}, $\tilde{H}=i\sigma_2 H^*$, $L$ is the lepton doublet of the second generation and $y$ quantifies the strength of the Yukawa interaction. The first term is clearly a 5-dimensional operator and it can be easily UV completed with 
new physics at TeV-scale, as was shown in e.g. \cite{Brdar:2020quo}. The second term in \cref{eq:1} is Majorana mass term for sterile neutrino, $N$, and the last term is the usual lepton portal through which $U_{\mu 4}$ gets generated. Let us stress that in this work we do not aim to explain active neutrino mass via seesaw mechanism;
the last two terms in \cref{eq:1} are only to show the minimal scenario for realization of mixing between sterile and active states.  


\textbf{Neutrino-Induced Photon flux.}
Given the model, in principle there are two options for sterile neutrino production from pion/meson decays -- via mixing and through the transition magnetic moment portal. As already argued above, even in the presence of nonzero $d_\mu$, the production of sterile neutrinos  is expected to be mixing-dominated.
The production via magnetic moment portal is suppressed since we expect similar numbers of $\pi^\pm$ and $\pi^0$ produced; each $\pi^\pm$ almost always gives an active neutrino in the final state, while both $\pi^\pm$ and $\pi^0$ have $d_\mu^2$ and phase suppressed 3-body decay into $N$, making the ratio of sterile to active neutrinos emerging from GRBs via magnetic moment to be $\sim d^2_\mu m^2_\pi/4\pi$. The production via mixing is only $|U_{\mu 4}|^2$ suppressed and clearly dominates.

After being produced, sterile neutrinos will propagate and along the way decay to photons and active neutrinos.
The expression for such photon flux from an astrophysical source reads
\begin{align}
|U_{\mu 4}|^2 \bigg[\frac{d/\tau}{\gamma/\Gamma-d/\tau} \left(\exp[-\frac{d\Gamma}{\gamma}]-\exp[-\tau]  \right)   \bigg] \frac{dN_\nu}{dE}\,{\rm Br}_\gamma\,,
\label{eq:2}
\end{align}  
where $\tau$ is the optical depth for sources located at $z= 0.15$ (extracted from \cite{Franceschini:2017iwq}), $\Gamma$ is the total decay rate of sterile neutrino (${\rm Br}_\gamma$ is the respective branching ratio for decays to final states containing photons), $\gamma=E_N/m_N$, $(dN_\nu/dE)$ is taken at the value of previously quoted IceCube limit and note that in computations we assume $E_\gamma \sim E_\nu$. Finally, the expression in the square brackets quantifies the likelihood for propagation of $N$ and $\gamma$ from GRB source at distance $d$ to Earth and we find agreement with respective formulae in \cite{Smirnov:2022suv}.

Let us now specify the channels contributing to $\Gamma$. In the presence of mixing, $N$ decays into three neutrinos ($N\to 3\nu$) at tree-level or radiatively into a neutrino and a photon ($N\to \nu\gamma$). 
In the presence of transition magnetic moment, there is another $N\to \nu\gamma$ decay channel; note that we are safe to neglect the interference terms between mixing and magnetic moment in the context of the two-body ($N\to \nu\gamma$) channel. The relevant decay widths of $N$ in our model are given by \cite{Zatsepin:1978iy,Pal:1981rm,Magill:2018jla}
\begin{eqnarray}
    \Gamma^{(3, U_{\mu 4})}_N&\approx&\frac{G^2_F m^5_N}{64\pi^3}|U_{\mu 4}|^2\,,\\
    \Gamma^{(2, U_{\mu 4})}_N&\approx& \frac{9\alpha G^2_F m^5_N}{512\pi^4}|U_{\mu 4}|^2\,,\\
    \Gamma^{(2, d_\mu)}_N&\approx& \frac{|d_\mu|^2 m^3_N}{4\pi}\,.
\label{eq:3}
\end{eqnarray}

With non-zero values of $U_{\mu 4}$ and $d_\mu$, we have therefore two options for decay into photon. For a given $m_N$ and with $d_\mu=0$, radiative decay via mixing will always be suppressed with respect to the tree-level contribution, making ${\rm Br}_\gamma= \mathcal{O}(10^{-3})$ in \cref{eq:2}. Notice that decays via mixing and magnetic moment scale with different power of $m_N$; therefore for smaller masses one can be in situation where decay via magnetic moment dominates and in that case ${\rm Br}_\gamma\approx 1$. 

Given all of the above, we are about to investigate the following scenarios for sterile neutrino
\begin{itemize}
\item production and decay via mixing\,,
\item production via mixing and decay via transition magnetic moment portal\,,
\end{itemize}
and we will demonstrate that only the latter case leads to the explanation of LHAASO's high energy photons.\\

 
\textbf{Results.} In order to predict a number of 18 TeV photon events from sterile neutrino decay, 
one needs to convolute the flux in \cref{eq:2} with the LHAASO effective area given e.g. in \cite{Ma:2022aau}. We consider bin-width of 8 TeV centered at 14 TeV.

\begin{figure}[t]
	\centering
	\includegraphics[scale=0.45]{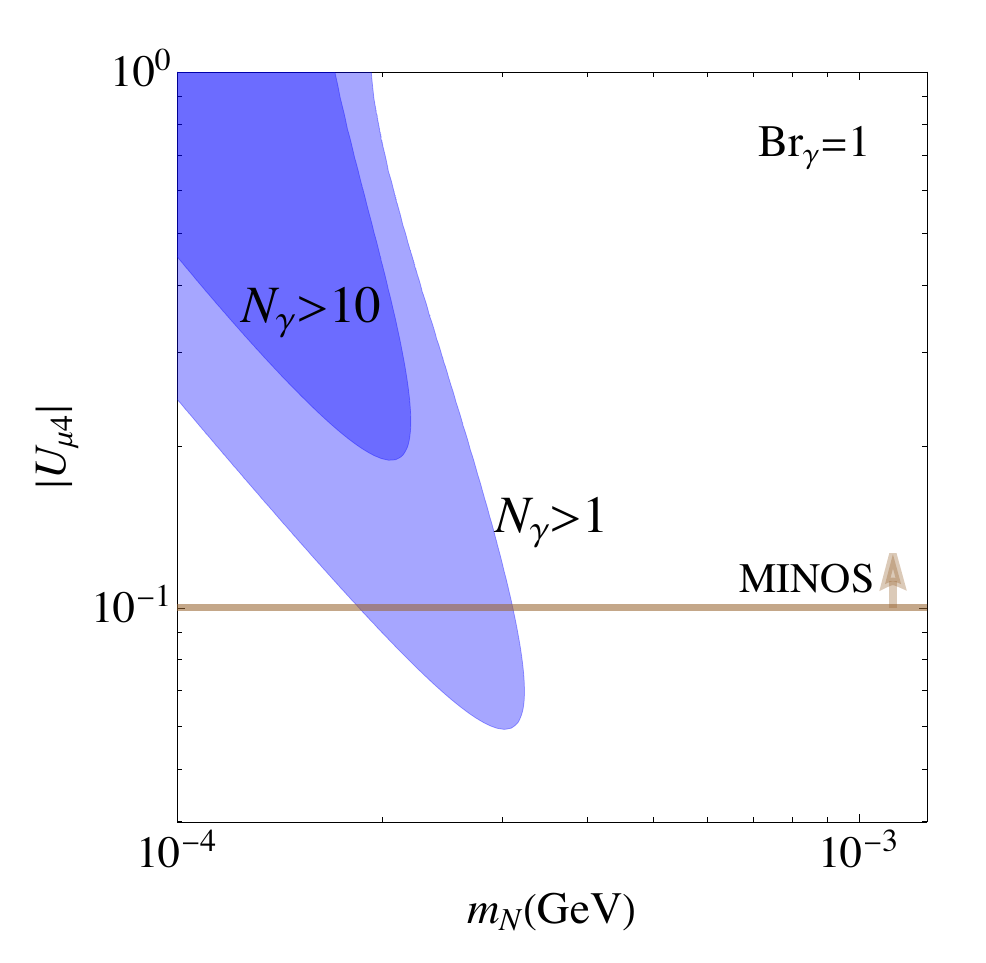} 
	\caption{Predicted number of high-energy events at LHAASO for the scenario in which both production and decay of sterile neutrino 
 occur via mixing. These regions are compared to the muon neutrino disappearance searches from MINOS. }
	\label{fig:mixing}
\end{figure}

Results for the mixing-only case are shown in \cref{fig:mixing}. Despite the discussed dominance of tree-level decay rate, we choose ${\rm Br}_\gamma=1$ in order to maximize the photon flux. The blue (light blue) region in the plot indicates the parameter space where more than 10 (1) events in the respective energy bin is obtained. We also checked that no events in higher bins as well as moderate number of low energy events (much less than LHAASO's reported $\mathcal{O}(5000)$) is obtained. The relevant sterile neutrino masses are $\mathcal{O}(\text{few}\times 10^{-1} \text{MeV})$ which is expected given sterile neutrino propagation length of $\approx d$. We compare our findings with the muon neutrino disappearance limit from MINOS that approximately reads  $|U_{\mu 4}|^2=0.01$ \cite{MINOS:2017cae}. 
While the constraints from non-unitarity (used in \cite{Smirnov:2022suv}) are somewhat stronger \cite{Blennow:2016jkn}, we argue that those are associated to heavy sterile neutrinos while MeV-scale sterile masses in our scenario are kinematically accessible at acceleration-based neutrino experiments. We nevertheless found that the minimal value of $|U_{\mu 4}|^2$ that would yield a single event in LHAASO is around $2.5\times 10^{-3}$. Values $|U_{\mu 4}|^2\sim 10^{-2}-10^{-3}$ are not constrained from other terrestrial experiments for the mixing of muon neutrino with sterile state \cite{Bolton:2019pcu}. However, this region is disfavored by standard cosmology and a richer dark sector \cite{Farzan:2019yvo,Cherry:2016jol,Chu:2018gxk} would need to be called upon. The successful explanation of 18 TeV event with neutrinos points to non-standard cosmology in which the bounds on the number of relativistic degrees of freedom can be satisfied.

Clearly, for both production and detection via mixing, a rather small parameter space is viable and we stress again that such statement holds only for ${\rm Br}_\gamma=1$ which is extremely optimistic value. Given the decay rates in mixing case, realistic ${\rm Br}_\gamma$ is $\mathcal{O}(10^{-3})$, and hence at most  $10^{-3}$ events are expected. Hence, the mixing scenario can not explain LHAASO anomaly and that conclusion was actually reached in \cite{Smirnov:2022suv} as well.

\begin{figure}[t]
	\centering
	\includegraphics[scale=0.45]{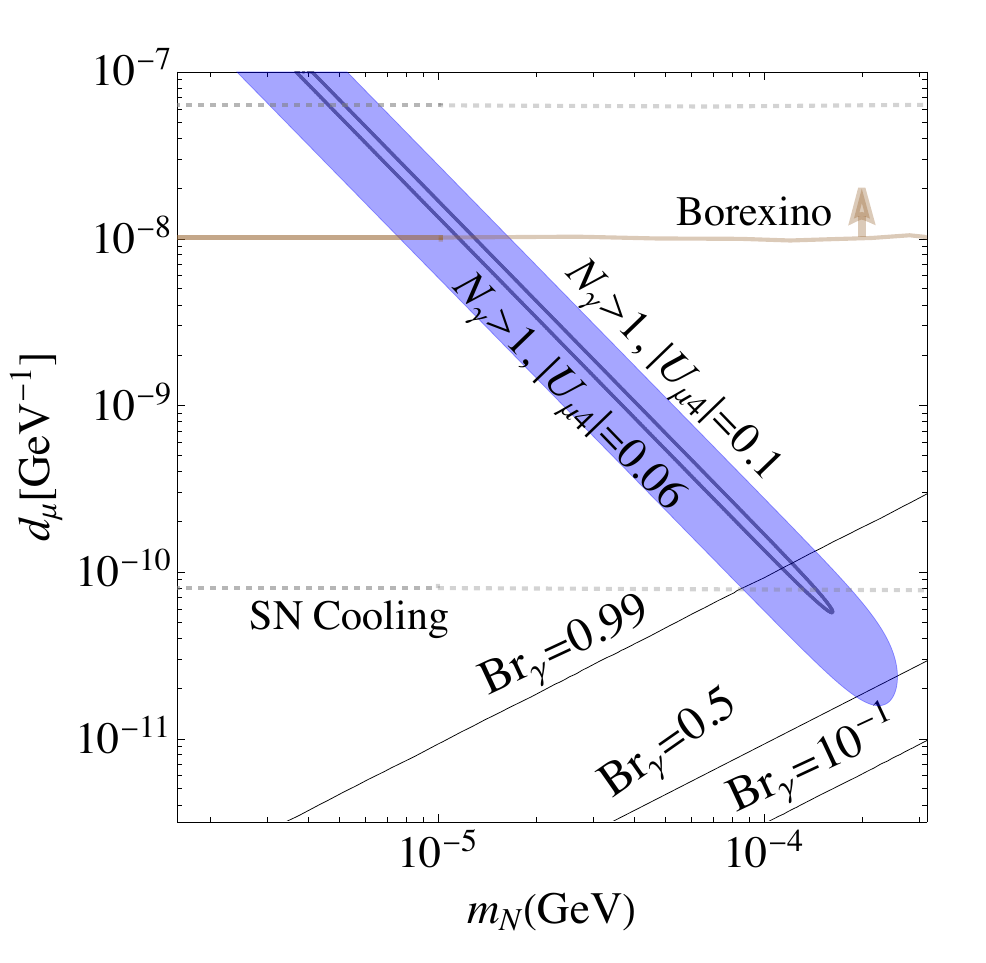} 
	\caption{Predicted number of high-energy events at LHAASO for the scenario in which production and decay of sterile neutrino 
 occur via mixing and magnetic moment portal, respectively.  We consider two values for  active-sterile mixing that are both consistent
 with constraints from oscillation experiments. We also show constraints on the magnetic moment portal from Borexino and cooling of Supernovae.}
	\label{fig:magnetic_mixng}
\end{figure}

Now, let us show results for nonzero $d_\mu$ where sterile neutrino decay goes via magnetic moment portal. As argued, this term will not alter production rate but it may significantly alter observational prospects as ${\rm Br}_\gamma \approx 1$ is achievable. In \cref{fig:magnetic_mixng}, in the $m_N-d_\mu$ parameter space we show again, in blue, regions in which 1 high-energy LHAASO event is achievable and we do that for two values of $U_{\mu 4}$. Notice that this region is now shifted to smaller $m_N$ with respect to 
the one in \cref{fig:mixing}. The reason for this is obvious and can be seen by just comparing decay rate formulae: the decay rate via magnetic moment operator (mixing) goes with $m_N^3$ ($m_N^5$) and, as a consequence, for smaller values of $m_N$,  the decay rate via magnetic moment operator would dominate over tree-level decay through mixing. We also have more freedom in $m_N$, when comparing to \cref{fig:mixing}; increase in the coupling can be compensated with a respective decrease of $m_N$.  The viable region in the parameter space for explaining 18 TeV event is chiefly immune to the most relevant bounds from terrestrial experiments (most importantly limit from neutrino-electron scattering in Borexino \cite{Brdar:2020quo}).  Astrophysical constraints from Supernovae \cite{Magill:2018jla} cooling cuts larger portion of the interesting parameter space; however, we should stress that such limits are still under debate \cite{Bar:2019ifz}. \\

We stress that in this model, given $m_N \lesssim 10^{-4}$ GeV, explanation of 18 TeV photon is possible due to $\mathcal{O}(1)$ value of ${\rm Br}_\gamma$. In \cref{fig:magnetic_mixng} we also demonstrate this by showing the ratio of decay rate via magnetic moment and via mixing. The region where we find 1 or more events is associated to ${\rm Br}_\gamma \gtrsim 1/2$.\\

Finally, let us comment on the time delay of this secondary photon flux with respect to the primary one, assuming that the active neutrino flux and primary photon flux are produced with negligible time difference. The delay occurs because of nonvanishing $m_N$ and we found $\Delta t\lesssim 50$ sec for all viable points in the parameter space; this falls well within LHAASO's 2000 sec observational window. Since we identified a large viable range in sterile neutrino mass, we can also accommodate a rather wide window of time delays of photons arising from sterile neutrino decays. This will be important upon LHAASO's upcoming more detailed characterization of the observed event.

\textbf{Summary and Conclusions.}
In this work we demonstrated that the minimal viable framework for explaining LHAASO's 18 TeV photon event with neutrinos includes both mixing and transition magnetic moment portal. In our setup, sterile neutrinos are produced at the source via mixing and they subsequently decay in the vicinity of Earth via transition magnetic moment portal. 
We identified the parameter space in which ${\rm Br}_\gamma\simeq 1$ for a wide mass range, leading to explanations of LHAASO's observation. Namely, we have found a substantial parameter space where at least one very high-energy event is expected and we checked consistency with constraints from terrestrial experiments. In addition to explaining LHAASO event, in this work we have presented a novel way of probing new physics coupling to neutrinos using astrophysical sources. \\

\textbf{Acknowledgements.}
We would like to thank Miguel Escudero and Joachim Kopp for useful discussions. Y.-Y. L is supported by the NSF of China through Grant No. 12047502.
\\\\
\textbf{Note added.}
While completing this manuscript, preprints by Cheung \cite{Cheung:2022luv} and Smirnov and Trautner \cite{Smirnov:2022suv}
on this topic appeared. As already pointed out in the manuscript, the explanation via magnetic moment only \cite{Cheung:2022luv}
is disfavored by strongly suppressed sterile neutrino production rate. For the case of mixing \cite{Smirnov:2022suv}, the situation 
is better on the production side, but since ${\rm Br}_\gamma \ll 1$, that kind of scenario also does not lead to a viable explanation
and only $10^{-3}$ events are predicted at most. The authors of \cite{Smirnov:2022suv} also pointed out scenarios where larger
${\rm Br}_\gamma$ could be achieved.

\bibliography{refs}

\end{document}